\documentstyle[12pt, epsf]{article}
\newcommand{\btab}{\begin{tabbing}}
\newcommand{\etab}{\end{tabbing}}

\newcommand{\beqn}{\begin{equation}}
\newcommand{\eeqn}{\end{equation}}
\newcommand{\barr}[1]{\begin{array}{#1}}
\newcommand{\earr}{\end{array}}
\newcommand{\beqna}{\begin{eqnarray}}
\newcommand{\eeqna}{\end{eqnarray}}
\newcommand{\btablec}{\begin{table} \begin{center}}
\newcommand{\etablec}{\end{center} \end{table}}
\newcommand{\lapprox}{\stackrel{<}{\scriptstyle \sim}}
\newcommand{\gapprox}{\stackrel{>}{\scriptstyle \sim}}

\newcommand{\gapproxeq}{\lower.7ex\hbox{$\;\stackrel{\textstyle>}{\sim}\;$}}
\newcommand{\plabel}[1]{\label{#1}}
\newcommand{\pbibitem}[1]{\bibitem{#1}}
\begin{document}
\title{
\begin{flushright} \small{hep-ph/9806233} \end{flushright} \vspace{0.6cm}  
\Large\bf Gluonic Excitations in Mesons}
\vskip 0.2 in
\author{Philip R. Page \thanks{\small \em E-mail:
prp@t5.lanl.gov. Invited talk presented at the ``Workshop on Production Properties and Interaction of Mesons'' (MESON'98), 30 May -- 2 June 1998, Cracow, Poland. To be published in Acta Physica Polonica.}\\
{\small \em T-5, MS-B283, Los Alamos 
National Laboratory, }\\ {\small \em  P.O. Box 1663, Los Alamos, NM 87545.}}
\date{}
\maketitle
\begin{abstract}{We report on some interesting recent theoretical and experimental advances on $J^{PC}$ exotics and hybrid mesons. These are the decay selection rules governing $J^{PC}$ exotic decay, the experimental evidence for a $J^{PC} = 1^{-+}$ exotic in $\eta\pi$ and $\rho\pi$, and the production of charmonium hybrids at forthcoming $B$--factories.}
\end{abstract}
\bigskip

In the last two years states beyond the predictions of the 
quark model have emerged in several $J^{PC}$ sectors, for example

\begin{itemize}

\item A preliminary study by VES confirms that there is evidence for two isovector
$0^{-+}$ states in the mass region $1.4 - 1.9$ GeV \cite{amelin97}. The parameters of the resonances are mass $1790\pm 6\pm 12$ MeV and width $225\pm 9\pm 15$ MeV for the well--known resonance $\pi(1800)$; and mass $1580\pm 43\pm 75$ MeV and width $450\pm 60\pm 100$ MeV for the new resonance. All known quark models predict
only one state in this mass region. Hence there is evidence for degrees of freedom
beyond simple $q\bar{q}$.

\item Perhaps most striking is the embarrassment of riches
for isovector states which are $J^{PC}=1^{-+}$ exotic, i.e. whose $J^{PC}$ quantum numbers cannot
be built from a fermion--antifermion operator (with derivatives). E852 reports a resonance with
mass $1370 \pm 16 ^{+50}_{-30}$ MeV and width $385\pm 40 ^{+65}_{-105}$ MeV \cite{bnletapi}; and Crystal Barrel a resonance with mass $1400\pm 20\pm 20$ MeV and a width of $310\pm 50 ^{+50}_{-30}$ MeV \cite{cbar}. Both collaborations observe the resonance in $\eta\pi$. A second, distinct, resonance has appeared in $\rho\pi$. E852 reports
a mass of $1593\pm 8$ MeV with a width of $168\pm 20$ MeV~\cite{bnl97}. There have also been claims of an enhancement in the high mass region $\gapprox 1.6$ GeV in $f_1\pi,\; a_0(980)\rho$ and $\eta^{'}\pi$ \cite{ryabchikov97}. Because the quantum numbers are exotic, neither state qualifies as a conventional meson, indicating evidence for the existence of degrees of freedom beyond $q\bar{q}$. 

\end{itemize}

The emergence of degrees of freedom beyond conventional mesons signals the end of the
hegemony of the quark model as a comprehensive description of bound states. 
Indeed, it is the beginning of the emergence of new states which indicate the
presence of dynamical glue in QCD, i.e. of gluonic excitations in mesons, also called ``hybrid mesons''.   
For the $J^{PC}=0^{-+},\; 1^{-+}$ values under consideration, there are $J^{PC}$ exotic and hybrid mesons
predicted to exist in lattice QCD \cite{lacock96,milc}. 

\section{Symmetrization Selection Rules}

Hybrid mesons have been shown to be bound states with a total decay width
of ${\cal O} (\frac{1}{N_c})$ in the large number of colours $N_c$ expansion of
QCD \cite{cohen} --  the same behaviour as for conventional mesons. 
This means that hybrids are expected to be observable states in 
experiment.

The decay widths of $J^{PC} = 0^{+-},\; 1^{-+},\; 2^{+-},\; \ldots$ exotic hybrid mesons,
four--quark states and glueballs have recently been realized to exhibit decay 
selection rules to final state $J=0$ mesons, which have considerable generality in 
QCD \cite{page97sel1}.

Specifically, the decay width for $1^{-+}\rightarrow \eta\pi$ is believed to be tiny. 
This decay has attracted considerable study
and detailed arguments in QCD have been developed to show that
the connected part of the quenched Euclidean three point correlation function of the interpolating
fields for the three states vanish exactly, if isospin symmetry is assumed \cite{pene}. 
Connecting the three point correlators
to decay amplitudes encounters difficulties with the way limits are taken for
time $t\rightarrow\infty$ \cite{testa} which was not fully appreciated in ref. \cite{pene}. 
These effects can be grouped under the name ``final state interactions'', and such
an effect is indicated in Fig. 1. Final state interactions break the 
symmetrization selection rules. The contribution from the graph in Fig. 1 has been
estimated in a model to give an $\eta\pi$ width to a 1.6 GeV $1^{-+}$ hybrid of less than
$57\pm 14$ MeV \cite{don98}.

\begin{figure}[b]
\begin{center}
\leavevmode
\hbox{\epsfxsize=5 in}
\epsfbox{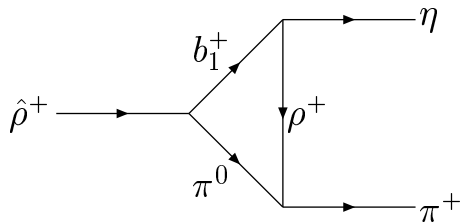}
\caption{\plabel{fig1} Decay of $1^{-+}$ to $\eta\pi$ via final state interactions.}
\end{center}
\end{figure}

\section{$1^{-+}$ exotics}

Some brief highlights from the history of the enhancement at 1.4 GeV in $\eta\pi$ is now presented. It was initially claimed that the enhancement is non--resonant and can be fitted with a Breit--Wigner mass of $1413$ MeV and a substantial width of $687$ MeV \cite{bugg94}. The Crystal Barrel recently claimed that the phase motion in the $\eta\pi$ P--wave is $213^o \pm 5^o$ and that the enhancement should hence be interpreted as resonant \cite{cbar}. E852 independently claimed that strong phase motion is observed against the $a_2$, indicating the resonant nature of the enhancement. 

Lattice QCD mass predictions and the predicted weak coupling of the $1^{-+}$ to $\eta\pi$ does not appear consistent
with the claim that the $1^{-+}$ at 1.4 GeV is indeed a hybrid meson \cite{page97exo}. We now explore a conservative
explanation for the 1.4 GeV enhancement.

The fact that non--trivial phase motion occurs is supported by both Crystal Barrel and E852, in very different
production processes. However, it is possible that the phase motion is really due to a resonant $1^{-+}$ at 1.6 GeV (where the $\rho\pi$ data favours it to be) interfering with a non--resonant background $\eta\pi$ P--wave, which ``shifts'' the peak from 1.6 GeV to 1.4 GeV. A recent K--matrix analysis has demonstated that this in indeed a strong possibility \cite{don98}.

The $1^{-+}$ at 1.6 GeV shows classic phase motion against the $a_2, \; a_1,\; \pi(1300), \; \pi(1800)$ and the $\pi_2(1670)$ \cite{bnl97}, so that its resonance nature is not in doubt. It is interesting to consider the various interpretations of a $J^{PC}$ exotic resonance in the general context of the minimally supersymmetric Standard Model. Here we think of building the bound states non--relativistically in terms of their perturbative constituents. Denote the gluon and gluino by $g$ and $\tilde{g}$ respectively; and the quark and squark by $q$ and $\tilde{q}$ respectively.
In principle one can imagine the following low--lying bound states
with exotic $J^{PC}$: glueballs ($gg$), gluinoballs ($\tilde{g}\tilde{g}$) and squarkballs ($\tilde{q}\bar{\tilde{q}}$). At this stage we do not consider
(more massive) bound states with a higher number of constituents. Note that $q\bar{q}$ does
not yield exotic $J^{PC}$ and that glueballinos ($g\tilde{g}$) and baryons ($qqq$) have half--integral spin, and so they should
not be considered. Within isospin symmetry, isovector glueballs and gluinoballs are not allowed, since the gluon and
gluino have isospin 0. If isospin symmetry is not assumed, the $\eta\pi$ can couple to an isoscalar glueball or
gluinoball. However, lattice QCD expects the $1^{-+}$ in the mass region $3-4.1$ GeV \cite{page97exo}, distant from the
mass region we are interested in. Also, gluinoballs cannot be $J^{PC}$ exotic, as we now show. 

Because the gluinos
are Majorana fermions, their parity is the imaginary number $i$. Hence the intrinsic parity of $\tilde{g}\tilde{g}$
is $i^2 = -1$. Adding this to the angular momentum $L$ between the gluinos, we obtain the parity $P=(-1)^{L+1}$. Because the
gluinos are their own antiparticles, the charge conjugation $C=1$. Since the gluinos are fermions, we require that the
gluinos must be antisymmetric under exchange. Since the gluinos are both
in colour octet and combine to form a colour singlet via a delta--function, which is symmetric under exchange, we
require the remainder of the wave function to be antisymmetric under exchange. Exchange of the gluinos just gives
$(-1)^{L+S+1}$, where $S$ is the non--relativistic spin of the gluinoball. Hence, $L+S$ must be even. Using the
relations

\beqn
P=(-1)^{L+1}\hspace{1cm} C=1\hspace{1cm} L+S = \mbox{even} \hspace{1cm} \vec{J} = \vec{L}+\vec{S}
\eeqn
it is easy to demonstate that no $J^{PC}$ exotics (i.e. $0^{--},\; 0^{+-},\; 1^{-+},\;  2^{+-},\; 3^{-+}, \ldots$)
can be built. 

Squarkballs cannot be $J^{PC}$ exotics either, as we now demonstate. Since the intrinsic parity of the squark and antisquark
are the same, $P=(-1)^L$. Because charge conjugation takes the particle into the antiparticle, and the squarks are spinless,
$C=(-1)^L$. Using these relations, and $\vec{J} = \vec{L}$, one can show that no $J^{PC}$ exotics are allowed.

Experimental evidence for isovector $J^{PC}$ exotic states in the $1-2$ GeV mass region hence indicates that they
are not conventional mesons, squarkballs, glueballs or glueballinos. The most conservative explanation are that they 
are hybrid mesons ($q\bar{q}g$), or four--quark states ($qq\bar{q}\bar{q}$) or linear combinations. The 1.6 GeV $1^{-+}$ appears to be
consistent with expectations for a hybrid meson \cite{page97exo}.

If the hybrid lies at 1.6 GeV, an interesting observation can be made. The heavy quark expansion of QCD 
in Coulomb gauge \cite{swanson97} demonstrates that spin--orbit splittings of low--lying hybrids 
should be be as follows: $0^{-+} < 1^{-+} < 2^{-+}$ and $0^{+-} < 1^{+-} < 2^{+-}$. The ordering from $0$ to $2$ with
$1$ in between is required by the lowest order in the heavy quark expansion, and must be the same for
both sets \cite{swanson97}. Either $0$ or $2$ is low--lying. The stated ordering follows from the fact that $0^{+-} < 2^{+-}$ is the ordering found in lattice QCD \cite{lacock96}. If spin--orbit splittings
for heavy quark hybrids are a guide for light quark hybrids, as is the case for conventional mesons, 
we expect the $0^{-+} < 1^{-+}$ at  $1593\pm 8$ MeV \cite{bnl97}. The new $0^{-+}$ at $1580\pm 43\pm 75$ MeV found by VES \cite{amelin97} may or may not satisfy this bound.  $\pi(1800)$ does not satisfy the bound. Since lattice QCD also supports that $1^{-+} < 0^{+-}$ 
\cite{lacock96,page97exo}, we can deduce the ordering $0^{-+} < 1^{-+} < 0^{+-} < 1^{+-} < 2^{+-}$ 
in the heavy quark expansion. It may in fact not be coincidental that non--$Q\bar{Q}$ degrees of freedom are
appearing in the two low--lying $J^{PC}$ combinations in this list. 
Whether the heavy quark limit is successful or not for light hybrids
will indicate whether there can be a successful ``quark model'' for hybrids.

\section{Production of Charmonium Hybrids at $B$--factories}

The weak reaction $b\rightarrow c\bar{c}s$ where the $c\bar{c}$ is in a colour octet is enhanced 
by colour factors above colour singlet $c\bar{c}$ production \cite{dunietz97}.
In the weak decay vertex the $c\bar{c}$ in the hybrid will be in a colour octet
at small interquark seperations,  as in the adiabatic
bag model, which is very successful when compared with lattice QCD \cite{morningstar97bag}. We hence expect 
$B$ mesons to decay significantly to $c\bar{c}$ hybrids ($\psi_g$).
Estimates in NRQCD indicate that ${\cal B}(B\rightarrow 0^{+-}\;  X) \lapprox \frac{1}{2} 0.1\%$ \cite{petrov98},
so that the branching ratio to all hybrids can be ${\cal O} (1\%)$.
 Given that CLEO has already detected
the $\chi_{c2}$ with a tiny branching ratio of ${\cal B}(B\rightarrow \chi_{c2} \;  X) = 0.25\pm 0.10 \%$,
there appears to be no reason why a search for charmonium hybrids should not be feasable
at B--factories.  
%The branching ratio of $B$ to charmonium is constrained by various experimental measurements.
%If the $B$ decay branching ratio to the low--lying hybrids ${\cal B}(B\rightarrow (\psi_g \mbox{ of all } J^{PC})\;  X) = {\cal O} (1\%)$, then the branching ratio of hybrids to charmonium
%${\cal B}(\psi_g\rightarrow c\bar{c}\; X) \leq {\cal O} (10-100\%)$ is consistent with the constraints from 
%experiment \cite{dunietz97}. 
%We shall see below that ${\cal B}(\psi_g\rightarrow c\bar{c}\; X)$ is indeed
%expected to be consistent with the experimental contraints. 

UKQCD's quenched lattice QCD calculation with infinitely heavy quarks predicts the low--lying hybrids, including the 
$1^{-+}$ and $0^{-+}$, to be at $4.04\pm 0.03$ GeV (with unquenching estimated to raise the mass by 0.15 GeV),
well below the $D^{\ast\ast}D$ threshold of 4.27 GeV \cite{dunietz97}.
We  highlight two $J^{PC}$ exotic hybrids of specific interest.

\vspace{.4cm}

\underline{$1^{-+}$}: 

\vspace{.2cm}

A quenched lattice QCD calculation by MILC predicts a mass of $4390\pm 80\pm 200$ MeV \cite{milc}. 
If the state
lies below the $D^{\ast\ast}D$ threshold, it will decay to $D^{\ast}\bar{D},\; D{\bar{D}}^{\ast}$ which is 
estimated in models at $3-4$ MeV and $D^{\ast} {\bar{D}}^{\ast}$ which vanishes in the same model
\cite{swanson98}. 

The prominent decays will be either by cascade
$\psi_g  \rightarrow gg\; + \; c\bar{c}$
or by annihilation  $\psi_g(C=+) \rightarrow  gg \rightarrow $ light
hadrons. These are at the same order in $\alpha_s$.  The decay $\psi_g
\rightarrow$ light hadrons is expected to be favoured at least for $C=+$ states
for the following reason. A measure of the relative importance of the cascade width compared to the
annihilation width may be provided by
$\Gamma (\psi^\prime \rightarrow \psi \mbox{ light hadrons}) = {\cal O}$(0.16 MeV) versus
$\Gamma
(\eta^\prime_c\rightarrow$ light hadrons) $\simeq
\Gamma(\eta_c\rightarrow$ light hadrons) $\times\; \Gamma^{ee}
(\psi^\prime)/\Gamma^{ee} (\psi) = {\cal O}(5 {\rm\;MeV})$.
The $\psi^\prime \rightarrow \psi  \mbox{ light hadrons}$  gives information about
$\psi^\prime \rightarrow \psi gg$, while $\eta^\prime_c\rightarrow$ light
hadrons informs about
$\eta^\prime_c\rightarrow gg$.  Both processes are ${\cal O}(\alpha_s^2)$ in
rate.  Those rates suggest what to expect for cascade and annihilation decays
of charmed hybrids. The rates of
$\psi_g \to (c\overline{c}) +$ light hadrons and
$\psi_g(C=+) \to  $ light hadrons are both down by one power in $\alpha_s$.
Ignoring differences in
wave function overlaps  we roughly estimate
$\Gamma(\psi_g \to (c\overline{c}) +$ light hadrons)
$= {\cal O}(0.5 {\rm \;MeV})$
and $\Gamma(\psi_g(C=+) \to  $light hadrons) $= {\cal O}(20 {\rm \;MeV})$
\cite{dunietz97}.

\vspace{.4cm}

\underline{$0^{+-}$}:

\vspace{.2cm}

The decay $0^{+-}$ to $D\bar{D},\; D^{\ast}\bar{D},\; D{\bar{D}}^{\ast},\; 
D^{\ast} {\bar{D}}^{\ast}$ is forbidden by quantum numbers. 
MILC  predicts a mass of $4610\pm 110\pm 200$ MeV \cite{milc}.
If the state lies below the $D^{\ast\ast}D$ threshold, its prominent decays will hence be by cascade
or annihilation.  

The light hadron production rate from $\psi_g$ decays with $C= -$ is expected
to be suppressed by one power of $\alpha_s$ with regards to
$\psi_g(C=+)$ decays.
Note that the production rate of conventional charmonia 
from either $\psi_g(C=+)$ or $\psi_g(C=-)$ decays is expected to be of the
same order in $\alpha_s$ and thus similar.
We estimate that
$\Gamma(\psi_g \to c\overline{c}\; +$ light hadrons)
$= {\cal O}(0.5 {\rm \;MeV})$
and $\Gamma(\psi_g(C=+) \to  $light hadrons) $= {\cal O}(5 {\rm \;MeV})$.
\vspace{.5cm}

A detailed list of search channels can be found in ref. \cite{dunietz97}.

In conclusion, even though experimental evidence for gluonic excitations
in mesons is already emerging, forthcoming $B$--factories have tantalizing prospects for exciting dynamical glue.

\vspace{.9cm}

\noindent {\bf Acknowledgements}

\vspace{.2cm}

Discussions with Carl E. Carlson on the quantum numbers of gluino-- and squarkballs are 
acknowledged.

\end{document}